\documentclass[preprint,prd,superscriptaddress]{revtex4}
\usepackage{amsmath,amssymb}
\usepackage{graphicx}
\usepackage{dcolumn}
\usepackage{bm}
\newcommand{\p}{\partial}
\newcommand{\pslash}{p\kern-1ex /}
\newcommand{\lslash}{l\kern-1ex /}
\newcommand{\kslash}{k\kern-1ex /}
\newcommand{\dslash}{\p\kern-1.2ex /}
\newcommand{\Dslash}{{\cal D}\kern-1.5ex /}
\newcommand{\Aslash}{A\kern-1.2ex /}

\newcommand{\tr}{{\rm tr}}
\newcommand{\Tr}{{\rm Tr}}
\newcommand{\re}{{\rm Re}}

\newcommand{\Dodwf}{\mathcal{D}}
\newcommand{\bea}{\begin{eqnarray}}
\newcommand{\eea}{\end{eqnarray}}

\newcommand{\EQ}{\hspace{-2mm} &=& \hspace{-2mm}}

\newcommand{\BAN}{\begin{eqnarray*}}
\newcommand{\EAN}{\end{eqnarray*}}
\begin{document}

\newcommand{\NTU}{
  Physics Department, National Taiwan University, Taipei~10617, Taiwan  
}

\newcommand{\CQSE}{
  Center for Quantum Science and Engineering, National Taiwan University, Taipei~10617, Taiwan  
}

\newcommand{\CTS}{
  Center for Theoretical Sciences, National Taiwan University, Taipei~10617, Taiwan  
}

\newcommand{\RCAS}{
  Research Center for Applied Sciences, Academia Sinica,
  Taipei~115, Taiwan
}

\newcommand{\NIU}{
  Center for Science Education, National Ilan Univerisity, I-Lan 260, Taiwan
}

\preprint{NTUTH-14-505B}
 
\title{Decay Constants of Pseudoscalar $D$-mesons in Lattice QCD with Domain-Wall Fermion}

\author{Wen-Ping~Chen}
\affiliation{\NTU}

\author{Yu-Chih~Chen}
\affiliation{\NTU}

\author{Ting-Wai~Chiu}
\affiliation{\NTU}
\affiliation{\CQSE}

\author{Han-Yi~Chou}
\affiliation{\NTU}

\author{Tian-Shin~Guu} 
\affiliation{\NIU}

\author{Tung-Han~Hsieh}
\affiliation{\RCAS}

\collaboration{TWQCD Collaboration}
\noaffiliation


\begin{abstract}

We present the first study of the masses and decay constants of the pseudoscalar $ D $ mesons 
in two flavors lattice QCD with domain-wall fermion. 
The gauge ensembles are generated on the $24^3 \times 48 $ lattice with the extent $ N_s = 16 $ in the fifth dimension,  
and the plaquette gauge action at $ \beta = 6.10 $, for three sea-quark masses with corresponding  
pion masses in the range $260-475$ MeV. 
We compute the point-to-point quark propagators, and measure the time-correlation functions of 
the pseudoscalar and vector mesons. The inverse lattice spacing 
is determined by the Wilson flow, while the strange and the charm quark masses by 
the masses of the vector mesons $ \phi(1020) $ and $ J/\psi(3097) $ respectively. 
Using heavy meson chiral perturbation theory (HMChPT) to extrapolate to the physical pion mass, 
we obtain $ f_D = 202.3(2.2)(2.6) $ MeV and $ f_{D_s} = 258.7(1.1)(2.9) $ MeV. 

\end{abstract}

\maketitle


In the Standard Model (SM), the quark and antiquark of a charged pseudoscalar meson 
$ P $ (with quark content $ Q \bar q $) can decay into a charged lepton and 
its associated neutrino through a virtual $W$ boson.
This is the purely leptonic decay of the charged pseudoscalar meson.  
To the lowest order, the purely leptonic decay width can be written as 
\bea
\label{eq:decay_width}
\Gamma(P \to l \nu ) = \frac{G_F^2}{8 \pi} |V_{Qq}|^2 m_l^2 \left(1-\frac{m_l^2}{M_P^2} \right)^2 M_P f_P^2, 
\eea
where $ G_F $ is the Fermi coupling constant, $ V_{Qq} $ is the Cabibbo-Kobayashi-Maskawa (CKM) matrix element, 
$ m_l $ is the mass of the lepton, $ M_P $ is the mass of the charged pseudoscalar meson, and 
$ f_P $ is the decay constant of the charged pseudocalar meson, which is defined by the matrix element 
of the axial vector current between the QCD vacuum and the one-particle state of the charged pseudoscalar meson, 
\bea
\label{eq:fP_def}
\left<0| \bar q \gamma_\mu \gamma_5 Q (0) | P(\vec{q}) \right> = i f_P q_\mu. 
\eea
According to (\ref{eq:decay_width}), experimental measurement of the leptonic decay width 
gives a determination of the product $ |V_{Qq} | f_P $. 
If the value of $ f_P $ can be obtained from another experimental measurement, then
the value of $ | V_{Qq} | $ can be determined, which is crucial for testing the SM
via the unitarity of the CKM matrix, as a constraint for any new physics beyond the SM. 
On the other hand, if the value of $ f_P $ is unavailable from other experiments, 
then it must be determined theoretically from the first principles of QCD, 
before the value of $ |V_{Qq}| $ can be fixed.  
 
Theoretically, lattice QCD is a viable framework to tackle QCD nonperturbatively
from the first principles of QCD, by discretizing the continuum space-time on a 
4-dimensional space-time lattice \cite{Wilson:1974sk},
and computing physical observables by Monte Carlo simulation \cite{Creutz:1980zw}.
However, in practice, any lattice QCD calculation suffers from the discretization and finite volume errors, 
plus the systematic error due to the unphysically heavy $ u/d $ quark masses (with $ M_\pi > 140 $ MeV).    
Moreover, since all quarks in QCD are excitations of Dirac fermion fields, 
it is vital to preserve all salient features of the Dirac fermion field on the lattice, 
in particular, the chiral symmetry of the massless Dirac fermion field.
It is nontrivial to formulate Dirac fermion field with exact chiral symmetry at finite lattice spacing. 
This is realized through the domain-wall fermion (DWF) on the 5-dimensional lattice \cite{Kaplan:1992bt}
and the overlap-Dirac fermion on the 4-dimensional lattice \cite{Neuberger:1997fp}. 

In June 2005, we determined the masses and decay constants of pseudoscalar mesons $ D $ and $ D_s $ 
in quenched lattice QCD with exact chiral symmetry \cite{Chiu:2005ue},     
before the CLEO Collaboration announced their high-statistics measurement 
of $ f_D $ in July 2005. Our theoretical predictions of $ f_D $ and $ f_{D_s} $
turned out in good agreement with the experimental values from the CLEO Collaboration \cite{Artuso:2005ym,Artuso:2007zg}.
In 2007, we extended our study to the $B$-mesons \cite{Chiu:2007km}, and determined the masses and decay constants 
of $ B_s $ and $ B_c $, as well as the lowest-lying spectra of heavy mesons with quark contents  
$ b \bar b $, $ c \bar b $, $ s \bar b $, and $ c \bar c $. 

To remove the systematic error due to the quenched approximation, 
it is necessary to simulate lattice QCD with dynamical quarks. 
For lattice QCD with exact chiral symmetry, the challenge is how to perform 
the hybrid Monte Carlo (HMC) simulation \cite{Duane:1987de}
such that the chiral symmetry is preserved at a high precision and all topological sectors are sampled ergodically. 


During 2011-2012, we demonstrated that it is feasible to perform large-scale dynamical QCD
simulations with the optimal domain-wall fermion (ODWF) \cite{Chiu:2002ir},  
which not only preserves the chiral symmetry to a good precision, but also samples all topological sectors ergodically.
To recap, we perform HMC simulations of two flavors QCD on the $ 16^3 \times 32 $ lattice (with lattice spacing $ a \sim 0.1 $~fm),
for eight sea-quark masses corresponding to the pion masses in the range 228-565 MeV.  
Our results of the topological susceptibility \cite{Chiu:2011dz},  
as well as the mass and decay constant of the pseudoscalar meson \cite{Chiu:2011bm},  
are all in good agreement with the sea-quark mass dependence predicted by the 
next-to-leading order (NLO) chiral perturbation theory (ChPT).
This asserts that the nonperturbative chiral dynamics of the sea-quarks 
are well under control in our HMC simulations.
In this paper, we perform HMC simulations of two flavors QCD with ODWF 
on the $ 24^3 \times 48 $ lattice (with lattice spacing $ a \sim 0.062 $~fm), 
with the purpose of studying the charm physics in lattice QCD with exact chiral symmetry.


In general, the 5-dimensional lattice Dirac operator of ODWF can be written as \cite{Chen:2012jya}
\bea
\label{eq:D_odwf}
[\Dodwf(m)]_{xx';ss'} &=&
  (\rho_s D_w + 1)_{xx'} \delta_{ss'}
 +(\sigma_s D_w - 1)_{xx'} L_{ss'},
\eea
where $ \rho_s = c \omega_s + d $, $ \sigma_s = c \omega_s - d $, and $ c $, $d$ are constants.
The indices $ x $ and $ x' $ denote the sites on the 4-dimensional space-time lattice,
and $ s $ and $ s' $ the indices in the fifth dimension, while  
the lattice spacing $ a $ and the Dirac and color indices have been suppressed.
The weights $ \{ \omega_s, s=1,\cdots, N_s \} $ along the fifth dimension are
fixed according to the formula derived in \cite{Chiu:2002ir} such
that the maximal chiral symmetry is attained.
Here $D_w$ is the standard Wilson Dirac operator plus a negative parameter $-m_0 \; (0 < m_0 < 2)$,
\begin{equation}
(D_w)_{xx'} = -\frac{1}{2} \sum_{\mu} \left[
  (1-\gamma_\mu)U_\mu(x)\delta_{x+\hat{\mu},x'}
 +(1+\gamma_\mu)U^\dagger_\mu(x')\delta_{x-\hat{\mu},x'} \right]
 + (4 - m_0),
\end{equation}
where $U_\mu(x)$ denotes the link variable pointing from $ x $ to $ x + \hat\mu $.
The operator $ L $ is independent of the gauge field, and it can be written as
\BAN
L = P_+ L_+ + P_- L_-, \quad P_\pm = (1\pm \gamma_5)/2,
\EAN
and
\BAN
\label{eq:L}
(L_+)_{ss'} = (L_-)_{s's}= \left\{
    \begin{array}{ll}
      - m \delta_{N_s,s'}, & s = 1, \\
      \delta_{s-1,s'}, & 1 < s \leq N_s,
    \end{array}\right.
\EAN
where $ N_s $ is the number of sites in the fifth dimension,
$ m \equiv r m_q $, $m_q $ is the bare quark mass, and $ r = 1/[2m_0(1-d m_0)] $.

Including the action of Pauli-Villars fields (with bare mass $ m_{PV} = 1/r $), 
the partition function of ODWF in a gauge background can be written as 
\bea
Z_{odwf} = \int [d \Psi] [ d \bar \Psi] [d \Phi] [d \Phi^\dagger]  
           \exp \left\{-\bar \Psi \Dodwf(m) \Psi - \Phi^\dagger \Dodwf(1) \Phi \right\},   
\eea
which can be integrated successively to obtain the fermion determinant of the effective 4-dimensional Dirac operator \cite{Chiu:2002ir}
\bea
Z_{odwf} = \det D(m_q), \hspace{5mm} D(m_q) = (D_c + m_q)(1+r D_c)^{-1}, 
\eea   
where 
\bea
\label{eq:odwf_4d}
\begin{aligned}
D_c &= \frac{1}{r} \frac{1+ \gamma_5 S_{opt}(H)}{1-\gamma_5 S_{opt}(H)},
\hspace{5mm}
S_{opt}(H) = \frac{1-\prod_{s=1}^{N_s} T_s}{1 + \prod_{s=1}^{N_s} T_s}, \\
T_s &= \frac{1-\omega_s H}{1+\omega_s H},
\hspace{5mm}
H = c H_w ( 1 + d \gamma_5 H_w)^{-1},
\hspace{5mm}
H_w = \gamma_5 D_w (-m_0).
\end{aligned}
\eea
Here $ S_{opt}(H) = H R_Z(H) $, where $ R_Z(H)$ is the Zolotarev optimal
rational approximation of $ (H^2)^{-1/2} $. 

For HMC simulation of lattice QCD with ODWF, it is crucial to 
perform the even-odd preconditioning on the ODWF operator (\ref{eq:D_odwf})
such that the condition number of the conjugate gradient is reduced 
and the memory consumption is halved.
Now, separating the even and the odd sites on the 4-dimensional space-time lattice,
(\ref{eq:D_odwf}) can be written as
\bea
\Dodwf(m_q) =
\begin{pmatrix}
4 - m_0 & D_w^{\text{EO}} \\
D_w^{\text{OE}} & 4 - m_0
\end{pmatrix}
[c\omega(1+L)+ d(1-L)] + (1-L)
=
\begin{pmatrix}
X & D_w^{\text{EO}} Y \\
D_w^{\text{OE}} Y & X
\end{pmatrix},
\label{eq:D_odwf_eo}
\eea
where
\begin{equation}
X \equiv (4 - m_0 )[ c \omega(1+ L)+d(1-L)] + (1-L), \quad Y \equiv c \omega (1+L) + d(1-L).
\end{equation}
We further rewrite it in a more symmetric form by defining
\begin{equation}
\label{eq:m5}
M_5\equiv \omega^{-1/2} YX^{-1} \omega^{1/2}
= \left\{(4-m_0) + \omega^{-1/2} \left[ c (1+L)(1-L)^{-1}  + d \omega^{-1} \right]^{-1} \omega^{-1/2} \right\}^{-1},
\end{equation}
and
\begin{equation}
S_1\equiv \omega^{-1/2} YX^{-1} = M_5 \omega^{-1/2},
\quad S_2\equiv Y^{-1} \omega^{1/2}.
\end{equation}
Then Eq.~(\ref{eq:D_odwf_eo}) becomes
\begin{equation}
\Dodwf(m_q) =
S_1^{-1}
\begin{pmatrix}
1 & M_5 D_w^{\text{EO}}  \\
M_5 D_w^{\text{OE}} & 1
\end{pmatrix}
S_2^{-1}
=S_1^{-1}
\begin{pmatrix}
1 & 0 \\
M_5 D_w^{\text{OE}} & 1
\end{pmatrix}
\begin{pmatrix}
1 & 0 \\
0 & C
\end{pmatrix}
\begin{pmatrix}
1 & M_5 D_w^{\text{EO}} \\
0 & 1
\end{pmatrix}
S_2^{-1},
\label{eq:D_odwf_decomp}
\end{equation}
where the Schur decomposition has been used in the last equality, with the Schur complement
\begin{equation}
\label{eq:C_def}
C \equiv 1 - M_5 D_w^{\text{OE}} M_5 D_w^{\text{EO}}.
\end{equation}
Since $ \det\Dodwf = \det S_1^{-1} \cdot \det C \cdot \det S_2^{-1} $, and
$ S_1 $ and $ S_2 $ do not depend on the gauge field, we can just use $ C $
for the Monte Carlo simulation. After including the Pauli-Villars fields (with $ m_{PV} = 1/r $),
the pseudofermion action for two-flavors QCD (in the isospin symmetry limit $ m_u = m_d $) can be written as
\bea
\label{eq:Spf}
S_{pf} = \phi^\dagger C_1^\dagger ( C C^\dagger)^{-1} C_1 \phi, \quad C_1 \equiv C(m = 1), 
\eea
where $ \phi $ and $ \phi^\dagger $ are complex scalar fields carrying the same quantum numbers
(color, spin) of the quark fields. Including the gluon fields, the partition function for 2 flavors QCD
can be written as
\bea
\label{eq:Z_nf2}
Z = \int[dU][d\phi^{\dag}][d\phi]\exp\left(-S_g[U]-\phi^\dagger C_1^\dagger ( C C^\dagger)^{-1} C_1 \phi \right),
\eea
where $S_g[U]$ is the lattice action for the gluon field. Here we use the plaquette gauge action
\bea
S_g[U]=\beta\sum_{plaq.}\left\{1-\frac{1}{3} \re \Tr (U_p) \right\}, \hspace{4mm} \beta=\frac{6}{g^{2}}.
\eea
Further details of our HMC simulations of two flavors QCD can be found in Ref. \cite{Chiu:2013aaa} 
and a forthcoming long paper. 


\begin{table}[htbp]
\caption{Basic parameters of the gauge ensembles:  
the sea-quark mass, the number of gauge configurations in each ensemble, 
the inverse lattice spacing, the residual mass, and the pion mass.}   
\begin{center}
\begin{tabular}{|c|ccccc|}
\hline
    Ensemble & $ m_{sea} a$ & $ N_{cfg} $ & $a^{-1}$[GeV] & $ m_{res} a$ & $M_\pi$[MeV] \\
\hline
    A  &  0.005 & 535 &  3.194(11)(12) &  0.000057(5) &  259(10)(11)   \\
    B  &  0.010 & 506 &  3.138(16)(12) &  0.000046(4) &  347(7)(6)   \\
    C  &  0.020 & 501 &  3.044(13)(11) &  0.000028(3) &  474(6)(5)   \\
\hline
\end{tabular}
\end{center}
\label{tab:info}
\end{table}


We perform the hybrid Monte Carlo simulation of two flavors QCD on the $24^3 \times 48$ lattice 
with the plaquette gauge action at $ \beta = 6/g^2 = 6.10$,
for three sea-quark masses ($ m_{sea} a = 0.005, 0.01, 0.02 $) with the corresponding 
pion masses in the range 260-475 MeV. 
For the quark part, we use ODWF with $ c = 1, d = 0 $ (i.e., $ H = H_w $), 
$ N_s = 16 $, and $ \lambda_{min}/\lambda_{max} = 0.05/6.2 $.
For each sea-quark mass,  
we generate the initial 400-500 trajectories on a Nvidia GPU card (with device memory larger than 4 GB). 
After discarding initial 300 trajectories for thermalization, we sample one configuration
every 5 trajectories, resulting 20-32 ``seed" configurations for each sea-quark mass. 
Then we use these seed configurations as the initial configurations for independent simulations on 20-32 GPUs.  
Each GPU generates 200-250 trajectories independently.  
Then we accumulate a total of 5000 trajectories for each sea-quark mass. 
From the saturation of the binning error of the plaquette, as well as
the evolution of the topological charge, 
we estimate the autocorrelation time to be around 10 trajectories. 
Thus we sample one configuration every 10 trajectories, 
and obtain $\sim 500$ configurations for each ensemble. 
The basic parameters of these three ensembles are summarized in Table \ref{tab:info}.

\begin{figure}[!htb]
\begin{center}
\begin{tabular}{@{}cccc@{}}
\includegraphics*[height=6.0cm,width=5.5cm,clip=true]{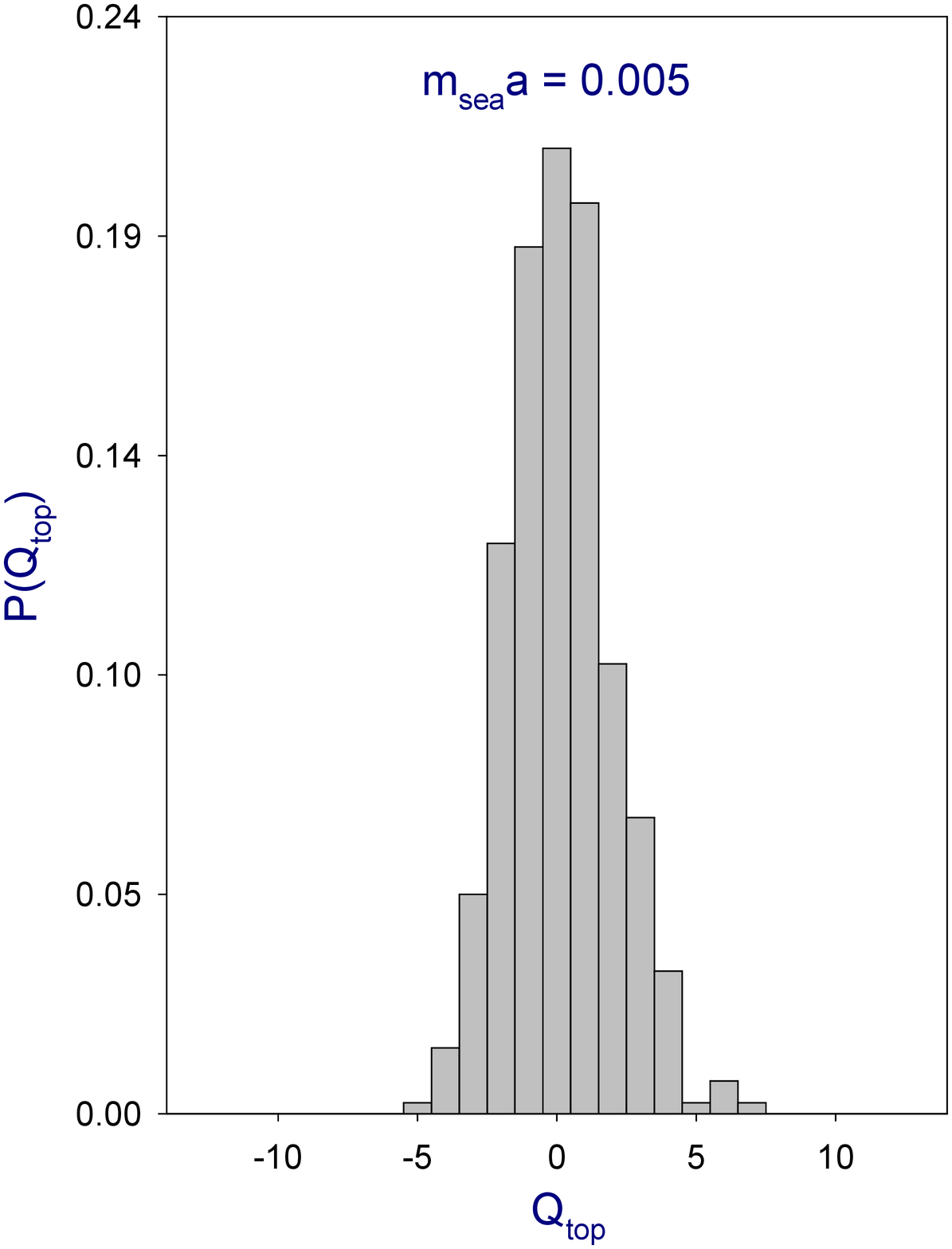}
&
\includegraphics*[height=6.0cm,width=4.5cm,clip=true]{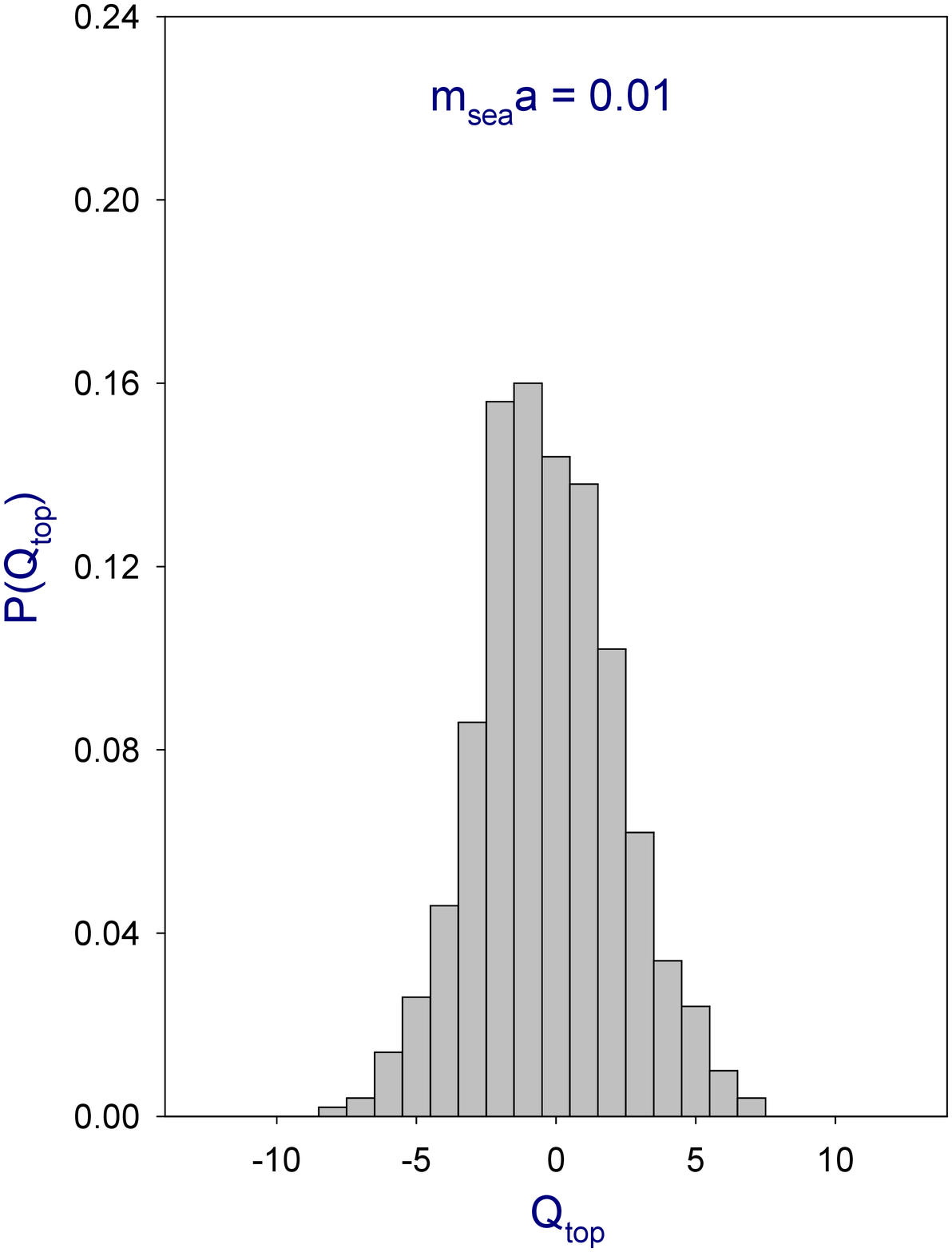}
&
\includegraphics*[height=6.0cm,width=4.5cm,clip=true]{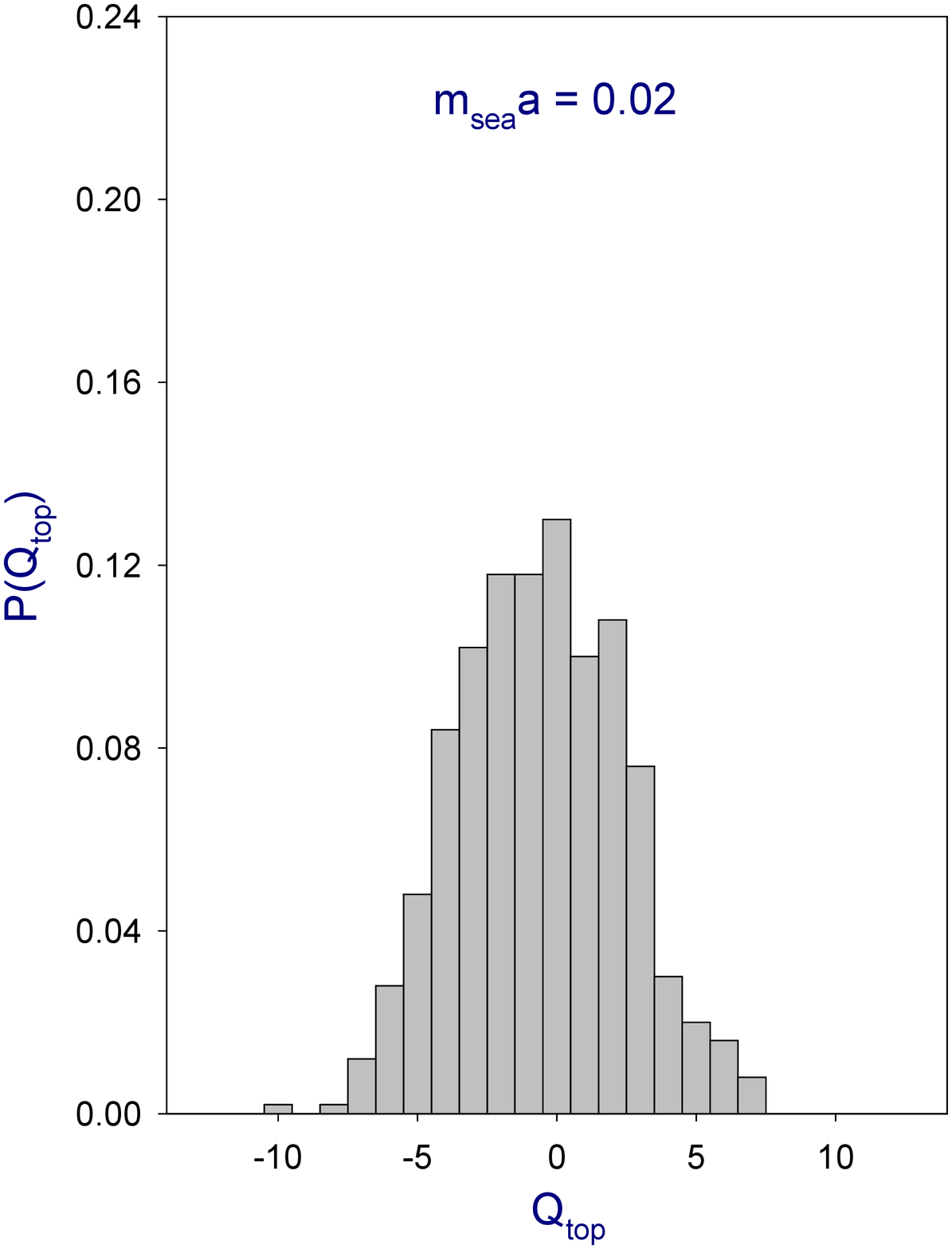}
&
\\
\end{tabular}
\caption{Histogram of topological charge distribution
for three sea-quark masses, $ m_{sea} a $ = 0.005, 0.01, and 0.02 respectively.
}
\label{fig:Q_hist}
\end{center}
\end{figure} 

In Fig.~\ref{fig:Q_hist}, we plot the histogram of the topological charge ($ Q_t $)   
distribution for these three ensembles. 
Evidently, the probability distribution of $ Q_t $ for each ensemble  
behaves like a Gaussian, and it becomes more sharply peaked around 
$ Q_t = 0 $ as the sea-quark mass gets smaller. 
Here the topological charge $ Q_t = \sum_x \epsilon_{\mu\nu\lambda\sigma} \tr[ F_{\mu\nu}(x) F_{\lambda\sigma}(x) ]/(32 \pi^2) $, 
where the matrix-valued field tensor $ F_{\mu\nu}(x) $ is obtained from the four plaquettes  
surrounding $ x $ on the ($\hat\mu,\hat\nu$) plane. Even though the resulting topological charge 
is not exactly equal to an integer, the probability distribution $ P(Q_t) $ suffices to demonstrate that 
our HMC simulation indeed samples all topological sectors ergodically.
For a rigorous study of topology, we are projecting the zero modes and the low-lying eigenmodes 
of the overlap Dirac operator for each gauge configuration, 
with the same procedures as outlined in Ref. \cite{Chiu:2011dz}, 
and we will report our results in a forthcoming long paper.     

To determine the lattice scale, we use the Wilson flow \cite{Narayanan:2006rf} with the condition \cite{Luscher:2010iy}
\bea
\label{eq:t0}
\left. \{ t^2 \langle E(t) \rangle \} \right|_{t=t_0} = 0.3,
\eea
to obtain $ \sqrt{t_0}/a $ for each gauge ensemble.
Our procedures are as follows. First, we compute the Wilson flow for each gauge ensemble of the 2-flavors QCD
on the $ 16^3 \times 32 $ lattice at $ \beta = 5.95 $ \cite{Chiu:2011bm}, and obtain the value of $ \sqrt{t_0}/a $. 
By linear extrapolation, we obtain $ \sqrt{t_0}/a = 1.3674(67)(42) $ in the chiral limit, 
where the systematic error is estimated by varying the number of sea-quark masses used for the linear extrapolation.
Then, using the lattice spacing $ a = 0.1034(1)(2) $~fm (in the chiral limit) \cite{Chiu:2011bm} 
which has been determined at $ \beta = 5.95 $ (by heavy quark potential with Sommer parameter $ r_0 = 0.49 $~fm),
we obtain $ \sqrt{t_0} = 0.1414(7)(5)$~fm. Thus, with the input $ \sqrt{t_0} = 0.1414(7)(5)$~fm, 
we can determine the lattice spacing of any gauge ensemble of 2-flavors QCD, with 
the value of $ \sqrt{t_0}/a $ obtained by the Wilson flow with the condition (\ref{eq:t0}). 
For the gauge ensembles of 2-flavors QCD on the $ 24^3 \times 48 $ lattice at $ \beta = 6.10 $, 
$a^{-1}$[GeV] = 3.194(11)(12), 3.138(16)(12), 3.044(13)(11),  
for $ m_{sea} a $ = 0.005, 0.01, and 0.02 respectively.
The inverse lattice spacing is well fitted by the linear function of $ m_{sea} a $, 
which gives $ a^{-1} = 3.2418(35)(44) $~GeV in the chiral limit. 

We compute the valence quark propagator with the point source at the origin, and with parameters exactly the same 
as those of the sea-quarks ($ N_s = 16 $ and $ \lambda_{min}/\lambda_{max} = 0.05/6.2 $).  
First, we solve the following linear system (with even-odd preconditioned CG), 
\bea
\label{eq:DY}
{\cal D}(m_q) |Y \rangle = {\cal D}(m_{PV}) B^{-1} |\mbox{source vector} \rangle, 
\eea
where $ B^{-1}_{x,s;x',s'} = \delta_{x,x'}(P_{-}\delta_{s,s'}+P_{+}\delta_{s+1,s'}) $
with periodic boundary conditions in the fifth dimension.
Then the solution of (\ref{eq:DY}) gives the valence quark propagator  
\BAN
\label{eq:v_quark}
(D_c + m_q)^{-1}_{x,x'} = r \left( 1 - r m_q \right)^{-1} \left[ (BY)_{x,1;x',1} - \delta_{x,x'} \right]. 
\EAN

For each gauge ensemble, we measure the time-correlation functions for pseudoscalar ($P$) and vector ($V$) mesons,
\BAN
\label{eq:CPS}
C_{P} (t) \EQ
\left<
\sum_{\vec{x}}
\tr\{ \gamma_5 (D_c + m_Q)^{-1}_{x,0} \gamma_5 (D_c + m_q)^{-1}_{0,x} \} \right>_U,   \\
\label{eq:CV}
C_V (t) \EQ \left<
\frac{1}{3} \sum_{\mu=1}^3 \sum_{\vec{x}}
\tr\{ \gamma_\mu (D_c + m_Q)^{-1}_{x,0} \gamma_\mu (D_c + m_q)^{-1}_{0,x} \} \right>_U, 
\EAN
for the following quark contents: 
$ (m_q, m_Q) $ = \{ $ (m_{sea}, m_{sea}) $, $(m_{sea}, m_{s})$, $(m_{sea}, m_{c})$, 
$(m_{s}, m_{s})$, $(m_{s}, m_{c})$, $(m_{c}, m_{c}) $ \}, where $ m_s $ and $ m_c $ are the bare masses 
of the strange quark and the charm quark. 



In general, the decay constant $ f_P $ of a charged pseudoscalar meson $ P $ with quark content $ Q \bar q $ 
is defined by (\ref{eq:fP_def}). In lattice QCD with exact chiral symmetry, we can 
use the axial Ward idenity 
\BAN
\partial_\mu  \left( \bar q \gamma_\mu \gamma_5 Q \right) = (m_q + m_Q) \bar q \gamma_5 Q,  
\EAN 
to obtain
\bea
\label{eq:fP}
f_P = (m_q + m_Q )
\frac{| \langle 0| \bar q \gamma_5 Q | P(\vec{0}) \rangle |}{M_P^2}, 
\eea
where the pseudoscalar mass $ M_P a $ and the decay amplitude
$ z \equiv | \langle 0| \bar q \gamma_5 Q | P(\vec{0}) \rangle | $
can be obtained by fitting the pseudoscalar time-correlation function
$ C_P(t) $ to the formula
\bea
\label{eq:Gt_fit}
\frac{z^2}{2 M_P a} [ e^{-M_P a t} + e^{- M_P a (T-t)} ], 
\eea
where the excited states have been neglected.

To measure the chiral symmetry breaking due to finite $N_s$, we compute the residual mass
according to the formula \cite{Chen:2012jya}
\bea
\label{eq:Mres}
m_{res}=\frac{\left< \tr(D_c + m_q)^{-1}_{0,0} \right>_U}{\left< \tr[\gamma_5 (D_c + m_q) \gamma_5 (D_c+m_q)]^{-1}_{0,0} \right>_U}-m_q,
\eea
where
$ (D_c + m_q)^{-1} $ denotes the valence quark propagator with $ m_q $ equal to the sea-quark mass, 
tr denotes the trace running over the color and Dirac indices, and the brackets $ \left< \cdots \right>_U $
denote averaging over an ensemble of gauge configurations. In Table \ref{tab:info}, we list the residual mass
of each ensemble. We see that the residual mass is at most $\sim 1$\% of the bare quark mass, amounting to 
$\sim 0.17 $ MeV, which is expected to be much smaller than other systematic errors.  
In Table \ref{tab:info}, we list the pion mass which is extracted from 
the time-correlation function $ C_P(t) $ with the valence quark masses equal to the sea-quark mass ($ m_q = m_Q = m_{sea} $).  
Here the pion mass has been corrected for the finite volume effect,  
using the estimate within ChPT calculated up to $ {\cal O}(M_\pi^4/(4 \pi f_\pi)^4 ) $ \cite{Colangelo:2005gd}. 

For each ensemble, the strange quark bare mass $ m_s $ is tuned such that 
the vector-meson mass $ M_V $ extracted from the time-correlation function 
$ C_V(t) $ with the valence quark masses $ m_q = m_Q = m_s $ is 
equal to the mass of the vector meson $ \phi(1020) $.  
The vector meson mass $ M_V $ is extracted by fitting $ C_V(t) $ to a formula similar to (\ref{eq:Gt_fit}),  
for the range $ [t_1, t_2] $ in which the effective mass attaining a plateau.  
We estimate the statistical error of $ M_{V} $  
using the jackknife method with the bin size (10-15 configuration) 
of which the statistical error saturates. To estimate the systematic errors of $ M_V $, 
besides that due to the systematic error of the inverse lattice spacing (see Table \ref{tab:info}),   
we also incorporate the variation of $ M_V $ based on all fittings satisfying $\chi^2$/dof $< 1.1$.  
In the following, it is understood that all masses and decay constants, and their errors 
are obtained with the same procedure.  
Similarly, the charm quark bare mass $ m_c $ is tuned such that 
the vector-meson mass extracted from the time-correlation function 
$ C_V(t) $ with the valence quark masses $ m_q = m_Q = m_c $ is 
equal to the mass of the vector meson $ J/\psi(3097) $. 
The values of $ m_s $ and $ m_c $ of each ensemble are summarized in 
Table \ref{tab:ms_mc}, where the error denotes the combined statistical and systematic error.   

\begin{table}[th]
\caption{Determination of the strange quark and the charm quark masses.}
\begin{center}
\begin{tabular}{|c|cccc|cccc|}
\hline
Ensemble & $ m_s a $ & $M_\phi$[MeV]     & $[t_1, t_2]$ & $\chi^2$/dof 
                     & $ m_c a $ & $M_{J/\psi}$[MeV] & $[t_1, t_2]$ & $\chi^2$/dof \\
\hline
A  & 0.04   & 1020(7)  & [11,24] & 0.99  & 0.53 & 3097(11) & [15,24] &  0.25  \\
B  & 0.04   & 1020(8)  & [11,23] & 0.11  & 0.55 & 3104(15) & [17,23] &  0.58  \\
C  & 0.04   & 1020(8)  & [16,23] & 0.20  & 0.55 & 3099(13) & [16,24] &  0.22  \\
\hline
\end{tabular}
\end{center}
\label{tab:ms_mc}
\end{table}


Using the value of $ m_c $ in Table \ref{tab:ms_mc}, 
we measure the time-correlation function of the $ D $ meson  
with the valence quark masses ($ m_q = m_{sea} $, $ m_Q = m_c $) 
for each ensemble.   
Then we fit the time-correlation function $ C_{D}(t) $ 
to (\ref{eq:Gt_fit}) to extract the $D$ meson mass $ M_D a $ 
and the decay constant $ f_D a $. 
Similarly, using the values of $ m_s $ and $ m_c $ in Table \ref{tab:ms_mc}, 
we measure the time-correlation function of the $ D_s $ meson  
with the valence quark masses ($ m_q = m_s $, $ m_Q = m_c $),  
and extract the mass $ M_{D_s} a $ and the decay 
constant $ f_{D_s} a $ for each ensemble. 
Our results of $ M_D $, $ f_D $, $ M_{D_s} $, and $ f_{D_s} $ 
are summarized in Table \ref{tab:fD_fDs}, 
where the error denotes the combined statistical and systematic error.   

\begin{table}[htbp]
\caption{The mass and decay constant of $D$-mesons.}   
\begin{center}
\begin{tabular}{|c|cccc|cccc|}
\hline
Ensemble & $M_D$[MeV] & $f_D$[MeV] & $[t_1,t_2]$ & $\chi^2$/dof 
         & $M_{D_s}$[MeV] & $f_{D_s}$[MeV] & $[t_1,t_2]$ & $\chi^2$/dof  \\
\hline
    A  & 1862(11) & 216(5) & [17, 24]  & 0.83  & 1969(8) & 261(5) & [14,23] & 0.97   \\
    B  & 1864(10) & 220(4) & [18, 23]  & 0.82  & 1968(9) & 264(4) & [13,24] & 0.54   \\
    C  & 1877(10) & 227(6) & [18, 22]  & 0.54  & 1962(8) & 267(3) & [10,17] & 0.70   \\
\hline
\end{tabular}
\end{center}
\label{tab:fD_fDs}
\end{table}

We see that for all ensembles, the masses of $ D $ and $ D_s $ mesons are 
in good agreement with the experimental values compiled by PDG \cite{Beringer:1900zz}, 
$ M_{D} = 1869.62 \pm 0.15 $~MeV, and $ M_{D_s} = 1968.50 \pm 0.32 $~MeV.

For the decay constants $ f_D $ and $ f_{D_s} $, we use 
HMChPT \cite{Sharpe:1995qp} to extrapolate to the physical $ M_\pi = 140 $~MeV.
In general, for the pseudoscalar meson with quark content ($ c \bar q $), the NLO formula for $ N_f = 2 $ reads
\bea
\label{eq:fDq_ChPT}
f_{D_q} = \frac{\kappa}{\sqrt{M_{D_q}}} 
          \left\{ 1-\frac{1+3 g_{c}^2}{2} \left[(\xi+\xi_q) \ln\left(\frac{\xi+\xi_q}{2} \right)-\frac{\xi_q}{2} \ln \xi_q \right] 
                  + c_1 \xi \right\},       
\eea
where $ \xi = M_\pi^2/(4 \pi f)^2 $,  $ \xi_q = M_{qq}^2/(4 \pi f)^2 $, 
$ M_{qq} $ is the mass of the pseudoscalar meson with quark content $ q \bar q $,  
$ g_c = 0.61(7) $ which is determined by the experimental measurement 
of the coupling $ g_{D^* \to D \pi} $ \cite{Anastassov:2001cw},  
and $ \kappa $ and $ c_1 $ are low-energy constants. 
For $ \bar q $ equal to $ \bar d $, (\ref{eq:fDq_ChPT}) reduces to 
\bea
\label{eq:fD_ChPT}
f_{D} = \frac{\kappa}{\sqrt{M_{D}}} \left[ 1-\frac{3}{4} (1+3 g_{c}^2) \xi \ln \xi + c_1 \xi \right].       
\eea

\begin{figure*}[tb]
\begin{center}
\begin{tabular}{@{}c@{}c@{}}
\includegraphics*[width=7.5cm,clip=true]{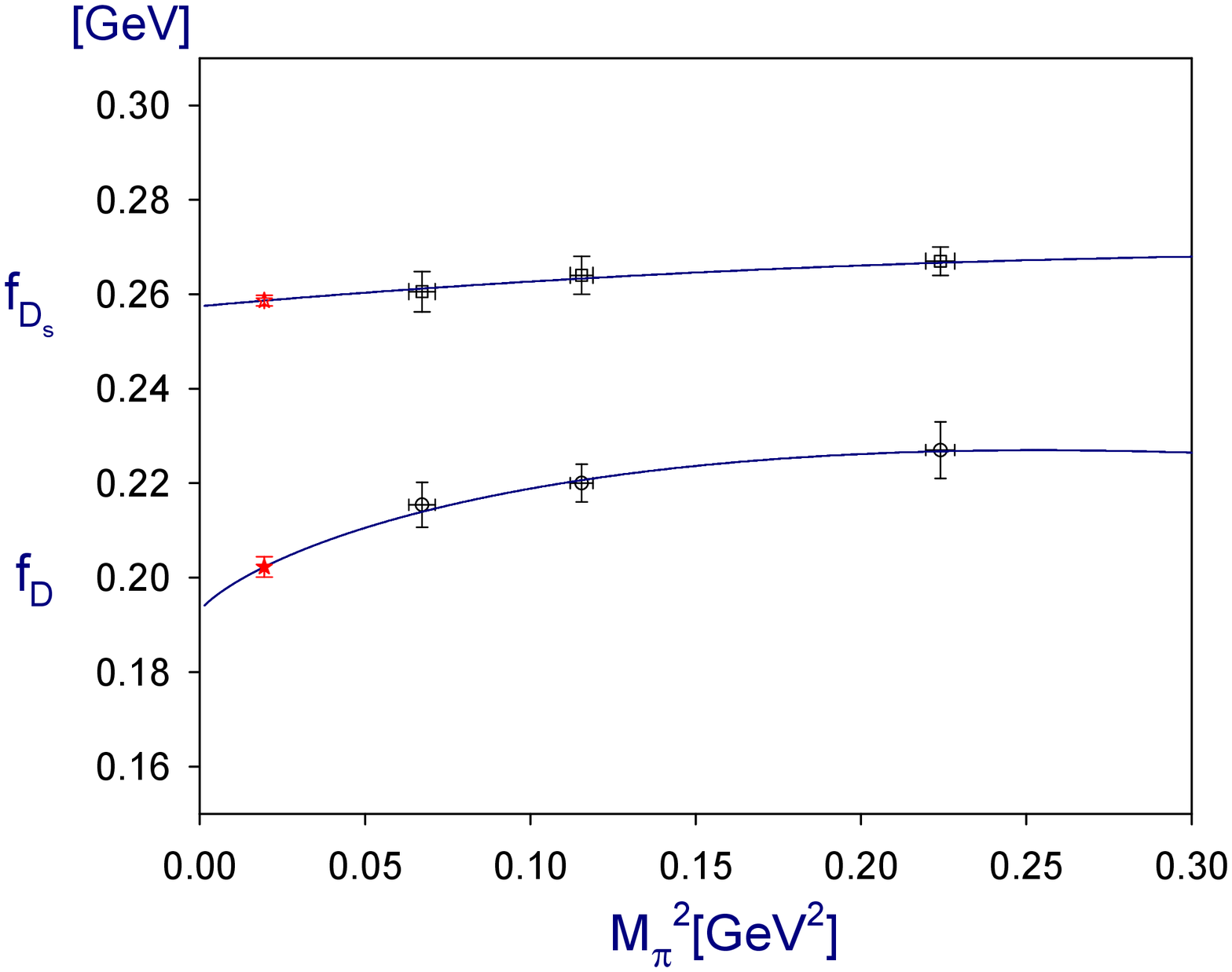}
&
\includegraphics*[width=7.5cm,clip=true]{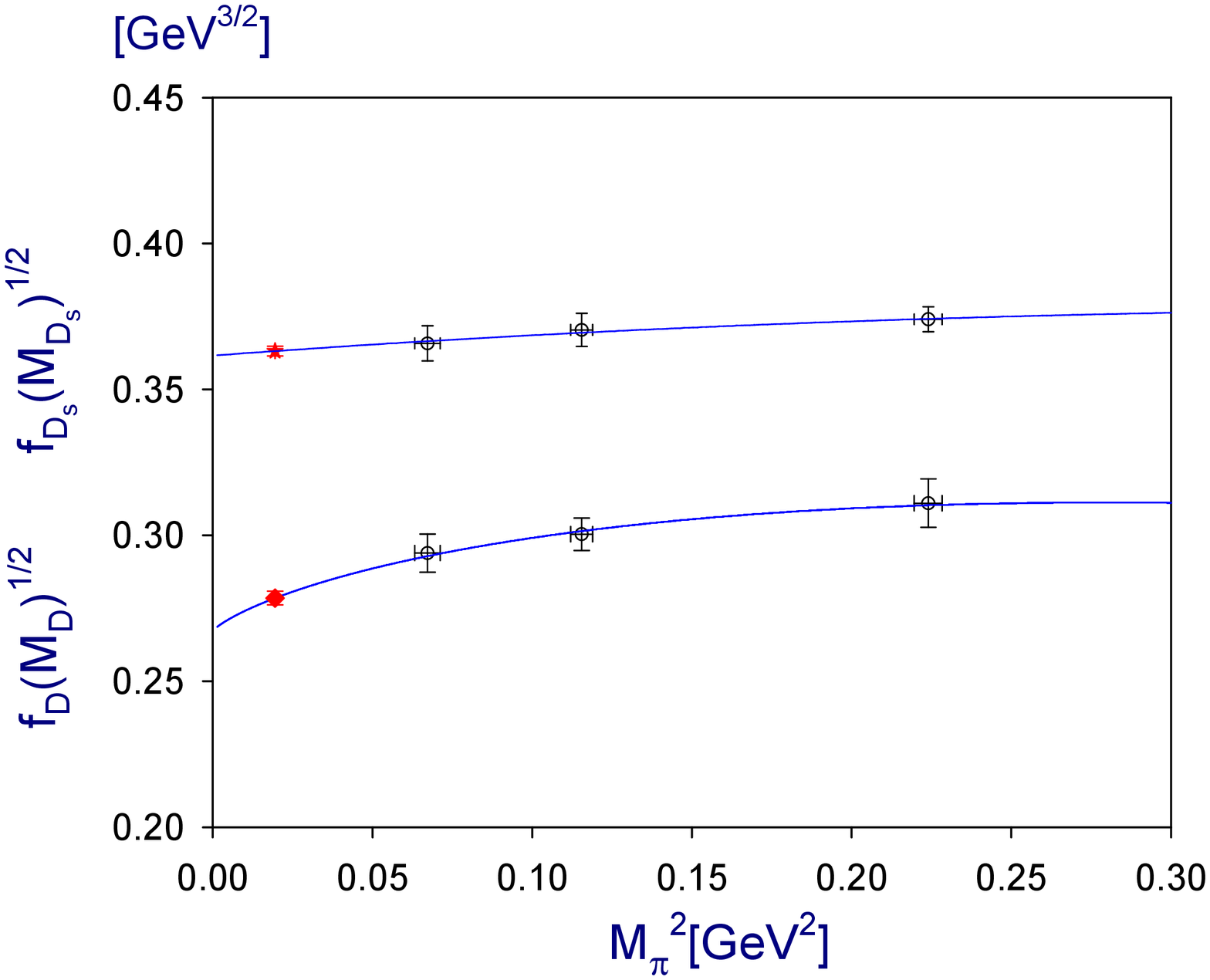}
\\ (a) & (b)
\end{tabular}
\caption{Fitting the data to NLO HMChPT: (a) $ f_D $ and $ f_{D_s} $; (b) $ f_D \sqrt{M_D} $ and $ f_{D_s} \sqrt{M_{D_s}} $.  
In each figure, the solid line is the fit of the data points to HMChPT. 
The symbol (in red) is the extrapolated value at physical $ M_\pi = 140 $ MeV.}  
\label{fig:fD_fDs}
\end{center}
\end{figure*}

\begin{figure*}[tb]
\begin{center}
\includegraphics*[width=7.5cm,clip=true]{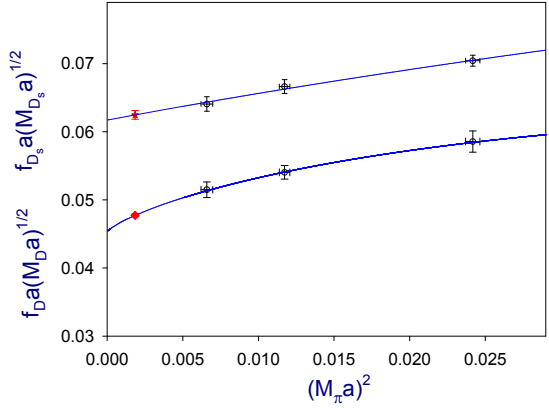}
\caption{ 
Fitting the lattice data of $ f_D a \sqrt{M_D a} $ and $ f_{D_s} a \sqrt{M_{D_s}a} $ (in the lattice unit) to NLO HMChPT. 
The solid line is the fit to HMChPT, and the symbol (in red) is the extrapolated value at physical $ M_\pi = 140 $ MeV.}  
\label{fig:rD_rDs}
\end{center}
\end{figure*}

Fitting the data of ensembles (A)-(C) to (\ref{eq:fD_ChPT}) 
[see the lower curve in Fig. \ref{fig:fD_fDs} (a)],   
we obtain $ \kappa = 0.2638(28) $ and $ c_1 = -1.899(206) $ with  $\chi^2$/dof = 0.34. 
At the physical $ M_\pi = 140$~MeV, (\ref{eq:fD_ChPT}) gives $ f_D = 202.3(2.2) $ MeV. 
Alternatively, performing the fit to $ f_D \sqrt{M_D} $ (multiplying both sides of (\ref{eq:fD_ChPT}) with $ \sqrt{M_D} $) 
[see the lower curve in Fig. \ref{fig:fD_fDs} (b)],  
we obtain $ \kappa = 0.2673(22) $ and $ c_1 = -2.012(189) $ with $\chi^2$/dof = 0.27. 
At the physical $ M_\pi = 140 $ MeV, it gives $ f_D \sqrt{M_D} = 0.2785(23) $, which in turn (with physical input $ M_D = 1869.62 $ MeV) 
yields $ f_D = 203.7(1.7) $ MeV. Comparing the results of these two HMChPT fits, we estimate the systematic error of $ f_D $ due to  
the chiral extrapolation to be $\sim 1.4$~MeV. 
Next we estimate the systematic error of $ f_D $ due to the scaling violations (i.e., the error induced through the scale setting),  
by performing the chiral extrapolation of $ (f_D a) \sqrt{M_D a} $ in the lattice unit,  
using the raw data of $ M_D a $ and $ f_D a $ extracted from the time-correlation function. 
Fitting the lattice data of ensembles (A)-(C) to NLO HMChPT (see the lower curve in Fig. \ref{fig:rD_rDs}) gives 
$\kappa = 0.0454(2) $ and $ c_1 = -0.6923(948) $ with $\chi^2$/dof = 0.14.
At the physical $ M_\pi = 140 $ MeV ($ M_\pi a = 0.001842 $), it gives $ f_D a \sqrt{M_D a} = 0.0477(2) $.
With the inputs $ a^{-1} = 3.2418(56) $~GeV (in the chiral limit) and $ M_D = 1869.62 $~MeV, we obtain $ f_D = 203.7(1.0) $~MeV.  
Thus we estimate the systematic error of $ f_D $ due to the scaling violations to be $\sim 1.0 $~MeV. 
Since our calculation is done at one single lattice spacing, 
the discretization error cannot be quantified reliably. 
Nevertheless, we do not expect it to be much larger than the systematic errors 
due to the chiral extrapolation and the scaling violations,   
beacuse our lattice action is free from $ O(a) $ discretization errors. 
Now, assuming the discretization error to be $\sim 2$~MeV, 
together with the systematic errors due to the scaling violations and the chiral extrapolation, we obtain  
\bea
\label{eq:fD}
f_D = 202.3 \pm 2.2 \pm 2.6~{\rm MeV},   
\eea 
which is in good agreement with the experimental values \cite{Eisenstein:2008aa,Huang:2012qc},  
as well as the experimental average $ f_D = 204.6 \pm  5.0 $ MeV \cite{Rosner:2013ica}.

Next, we turn to the decay constant of $ D_s $ meson. Fitting the data of ensembles (A)-(C) to (\ref{eq:fDq_ChPT}) with 
$ \bar q = \bar s $ [see the upper curve in Fig. \ref{fig:fD_fDs} (a)],   
we obtain $ \kappa = 0.2790(12) $ and $ c_1 = -0.7740(733) $ with $\chi^2$/dof = 0.20. 
At the physical $ M_\pi = 140$~MeV, (\ref{eq:fDq_ChPT}) gives $ f_{D_s} = 258.7(1.1) $~MeV.  
Alternatively, performing the fit to $ f_{D_s} \sqrt{M_{D_s}} $  [see the upper curve in Fig. \ref{fig:fD_fDs} (b)],  
we obtain $ \kappa = 0.2865(13) $ and $ c_1 = -0.8949(899) $ with $\chi^2$/dof = 0.21. 
At the physical $ M_\pi = 140 $ MeV, it gives $ f_{D_s} \sqrt{M_{D_s}} = 0.2785(23) $, 
which (with physical input $ M_{D_s} = 1968.50 $ MeV) yields $ f_{D_s} = 258.8(1.2) $ MeV. 
Comparing the results of these two HMChPT fits, we estimate the systematic error of $ f_{D_s} $ due to the  
chiral extrapolation to be $ \sim 0.2 $~MeV. 
Next, we estimate the systematic error of $ f_{D_s} $ due to the scaling violations,  
by performing the chiral extrapolation of $ (f_{D_s} a) \sqrt{ M_{D_s} a} $ in the lattice unit.  
Fitting the lattice data to NLO HMChPT (see the upper curve in Fig. \ref{fig:rD_rDs}) gives 
$\kappa = 0.0489(5) $ and $ c_1 = 0.513(185) $ with $\chi^2$/dof = 0.46.
At the physical $ M_\pi = 140 $ MeV ($ M_\pi a = 0.001842 $), it gives $ f_{D_s} a \sqrt{M_{D_s} a} = 0.0625(6) $.
With the inputs $ a^{-1} = 3.2418(56) $~GeV (in the chiral limit) and $ M_{D_s} = 1968.50 $~MeV, we obtain $ f_{D_s} = 259.9(2.7) $~MeV.  
Thus we estimate the systematic error of $ f_{D_s} $ due to the scaling violations to be $\sim 2.0 $~MeV. 
Again, assuming the discretization error to be $\sim 2$~MeV, 
together with the systematic errors due to the scaling violations and the chiral extrapolation, we obtain
\bea
\label{eq:fDs}
f_{D_s} = 258.7 \pm 1.1 \pm 2.9~{\rm MeV}, 
\eea 
which is in good agreement with the experimental values \cite{Artuso:2007zg,delAmoSanchez:2010jg,Zupanc:2013byn}, 
as well as the experimental average $ f_{D_s} = 257.5 \pm 4.6 $ MeV \cite{Rosner:2013ica}. 

Since the ratio $ f_{D_s}/f_D $ is expected to have systematic error 
less than those of $ f_{D} $ and $ f_{D_s} $, our results of $ f_D $ and $ f_{D_s} $ yields 
\bea
\label{eq:ratio}
\frac{f_{D_s}}{f_D} = 1.2788 \pm 0.0264, 
\eea   
in good agreement with the experimental average $ f_{D_s}/f_D = 1.258 \pm 0.038 $ \cite{Rosner:2013ica}. 

To summarize, we perform the first study of the masses and decay constants 
of the pseudoscalar $D$-mesons in two-flavors lattice QCD with domain-wall fermion,  
on the $24^3 \times 48 $ lattice with lattice spacing $ a \sim 0.062 $~fm, 
for three sea-quark masses corresponding to the pion masses in the range $260-475$~MeV. 
Our results of $ f_D $ (\ref{eq:fD}), $ f_{D_s} $ (\ref{eq:fDs}) and their ratio (\ref{eq:ratio}), 
are all in good agreement with the experimental results, as well as with other lattice QCD results \cite{Aoki:2013ldr}.
Since our calculation is done at one single lattice spacing, we cannot perform the extrapolation to the continuum limit.
Nevertheless, we do not expect the combined systematic errors much larger than our estimates in (\ref{eq:fD}) and (\ref{eq:fDs}), 
since the lattice spacing ($ a \sim 0.062 $~fm) is sufficiently fine, and our lattice action is free from $ O(a) $ lattice artifacts. 
Likewise, since our calculation is done on a single volume, the finite volume effect cannot be estimated reliably. 
However, it is believed that the finite volume error of physical observables involving heavy (charm/bottom) quarks  
is smaller than other systematic ones.  
We will address these issues with calculations on different volumes as well as several lattice spacings. 
Moreover, to incorporate the internal quark loops of ($u,d,s,c$) quarks in our dynamical simulations,  
we are performing HMC simulations of $(2+1)$-flavors and $(2+1+1)$-flavors QCD on the $ 24^3 \times 48 $ lattice,  
with the novel exact pseudofermion action for one-flavor DWF \cite{Chen:2014hyy}.

  This work is supported in part by the Ministry of Science and Technology 
  (Nos.~NSC102-2112-M-002-019-MY3,~NSC102-2112-M-001-011) and NTU-CQSE (No.~103R891404). 
  We also thank NCHC for providing facilities to perform part of our calculations.

\end{document}